\documentclass[pre,preprint,groupedaddress, showkeys,showpacs]{revtex4}

\usepackage{graphicx}
\begin{document}


\title{Heat conduction in deformable Frenkel-Kontorova lattices: thermal conductivity and negative differential thermal resistance}


\author{Bao-quan  Ai$^{1,2}$}
\author{Bambi Hu$^{2,3}$}

\affiliation{$^{1}$ Laboratory of Quantum Information Technology, ICMP and
 SPTE, South China Normal University, Guangzhou, China.\\
$^{2}$ Department of Physics, Centre for Nonlinear Studies, and the
Beijing-Hong Kong-Singapore Joint Centre for
Nonlinear and Complex Systems (Hong Kong), Hong Kong Baptist University, Kowloon Tong, Hong Kong, China\\
 $^{3}$Department of Physics, University of Houston, Houston, Texas 77204-5005, USA}


\date{\today}
\begin{abstract}
\indent Heat conduction through the Frenkel-Kontorova (FK) lattices
is numerically investigated in the presence of a deformable
substrate potential. It is found that the deformation of the
substrate potential has a strong influence on heat conduction. The
thermal conductivity as a function of the shape parameter is
nonmonotonic. The deformation can enhance thermal conductivity greatly and there exists an optimal deformable value at which thermal
conductivity takes its maximum. Remarkably, we also find that the
deformation can facilitate the appearance of the negative
differential thermal resistance (NDTR).
\end{abstract}




\maketitle
\section {Introduction}
\indent Heat conduction in low dimensional lattices is a well known
classical problem related to the microscopic foundation of
Fourier¡¯s law \cite{a1,a2}. The theoretical studies have not only
enriched our understanding of the microscopic physical mechanism of
heat conduction, but also suggested some useful thermal devices such
as rectifiers or diodes\cite{a3}, thermal transistors \cite{a4},
thermal logic gates \cite{a5}, and thermal memory \cite{a6}. People
would like to know whether or not the Fourier law of heat conduction
for bulk material is still valid in low dimensional systems. This is
a fundamental question in nonequilibrium statistical mechanics. In
general, the one-dimensional models can be classified into three
categories \cite{a7}. The first one is integrable system such as the
harmonic chain, in which no temperature gradient can formed and
thermal conductivity is divergent \cite{a8}. The second category
includes some nonintegrable systems such as the diatomic Toda chain
\cite{a9}, the Fermi-Pasta-Ulam (FPU) chain \cite{a10}, Heisenberg
spin chain\cite{a11}, and so on. In these models, the temperature
gradient can be formed but thermal conductivity is divergent at the
thermodynamics limit. The third category contains the nonintegrable
systems such as the ding-a-ling model \cite{a12}, Lorentz gas
model\cite{a13}, $\phi^{4}$ model\cite{a1}, and FK model
\cite{a14,a15}. In this category, thermal conductivity is finite and
the Fourier's law is justified.

\indent FK model was first proposed by Frenkel and Kontorova
\cite{a14} in 1938 to study surface phenomena. Since then it has
found application in wade variety of physics systems \cite{a15} such
as adsorbed monolayers, Josephson junctions, and DNA denaturation.
Despite its deceptively simple form, the model exhibits very rich
and complex behaviors in heat conduction. For example, one can
obtain the different frequency bands for different cases\cite{a3}:
$\sqrt{\frac{V}{mL^{2}}}<\omega<\sqrt{\frac{V}{mL^{2}}+4\frac{K}{m}}$ for low temperature limit, and
$0<\omega<2\sqrt{\frac{K}{m}}$ for high temperature limit, where $V$ is the
height of the substrate potential, $m$ is the mass of the particle, $K$ is coupling constant,
and $L$ is the period of the on-site potential. The dependence of thermal conductivity on the temperature, the strength
and the periodicity of the external potential, and the coupling
constant is extensive studied \cite{a7,a15}. Shao and co-workers
\cite{a16} studied the dependence of thermal conductivity $\kappa$
on the strength of the interparticle potential $K$ and the strength
of the external potential $V$ in the FK model and found the scaling
form $\kappa \propto \frac{K^{3/2}}{V^{2}}$. Barik \cite{a17}
studied the heat conduction of a two-dimensional FK lattice and
found anomalous heat conduction. Zhong \cite{a18} studied a
double-stranded system modeled by a FK lattice and found that the
interchain interaction has a positive effect on the thermal
conductivity in the case of strong nonlinear potential, and has a
negative effect on thermal conductivity in the case of weak
nonlinear potential. In our previous work \cite{a19}, we designed a thermal pumping
by applying an external ac driving force at one boundary of FK lattice and found that
the heat can be pumped from the low-temperature heat bath to the
high temperature one by suitably adjusting the frequency of the ac driving force.

\indent The study of heat conduction in low-dimensional systems
also has practical implications. It has recently been found that nonlinear systems with structural asymmetry can
exhibit thermal rectification, which has triggered model designs of
various types of thermal devices such as thermal transistors
\cite{a4}, thermal logic gates \cite{a5}, and thermal
memory\cite{a6}. It is worth pointing out that most of these
studies are relevant to heat conduction in the nonlinear response
regime, where the counterintuitive phenomenon of negative
differential thermal resistance (NDTR) may be observed and plays an
important role in the operation of those devices. NDTR refers to the
phenomenon where the resulting heat flux decreases as the applied temperature difference (or gradient)
increases. It can be seen that a comprehensive understanding of the
phenomenon of NDTR would be conducive to further developments in the designing and fabrication
of thermal devices.

\indent  Most of studies are involved the regular potentials.
However, in the real physical systems, the shape of the substrate
potential can deviate from the standard (sinusoidal) one, and this
may affect strongly the transport properties of the system
\cite{b1}. The effects of the shape of the substrate potential on
heat conduction is still lacking to date.
 In the present work, we study the heat conduction in deformable
FK chains by using nonequilibrium molecular dynamics simulations. We
emphasize on finding how the deformation of the potential affects
thermal conductivity and NDTR.

\section {Model and methods}
In this paper, we study heat conduction in FK chains with an
asymmetric deformable potential. The Hamiltonian of the whole system
is
\begin{equation}\label{}
    H=\sum
    _{i=1}^{N}\frac{p_{i}^{2}}{2m}+\frac{1}{2}K(x_{i}-x_{i+1})^{2}+U(x_{i}),
\end{equation}
where $p_{i}$ is the momentum of the $i$th particle, and $x_{i}$ its
displacement from equilibrium position. $m$ is the mass of the
particles, $K$ is the spring constant and $N$ is total number of the
particles. In order to investigate the effects of the shape of the
substrate potential on heat conduction, we use the following
asymmetric deformable potential (shown in Fig. 1)\cite{b1}:
\begin{equation}\label{}
    U(x)=\frac{V}{(2\pi)^{2}}\frac{(1-r^{2})^{2}[1-\cos(2\pi x)]}{[1+r^{2}+2r\cos(\pi
    x)]^{2}},
\end{equation}
where $V$ is the height of the potential and $r$ is the shape
parameter. The potential reduces to the simply sinusoidal
potential for $r=0$ or $|r|\rightarrow \infty$. The degree of the deformation will increase when
$|1-r^{2}|$ decreases. The potential will disappear for $|r|=1$.  For convenience of discussion,
we take the parameter range $0<r<1$,  and in this parameter range, the degree of the deformation increases with $r$.
For $0<r<1$, it is an asymmetric periodic one with a constant barrier height and two inequivalent successive wells
with a flat and sharp bottom, respectively. This potential is
considered as a natural way to describe lattice with diatomic basis
or dual lattices by generalizing the standard model that assumes
simple sinusoidal potential.
\begin{figure}[htbp]
\begin{center}\includegraphics[width=12cm,height=8cm]{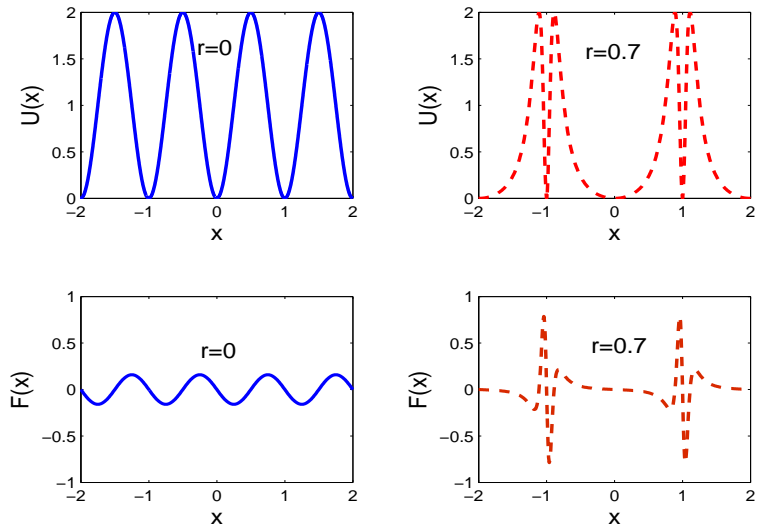}
\caption{(Color online) Asymmetric deformable potential for
different values of the shape parameter $r=0.0$ and $0.7$. The upper
shows the potential $U(x)$ described in Eq. (2) and the bottom
depicts its corresponding force $F(x)=-\frac{\partial U(x)}{\partial
x}$.}\label{1}
\end{center}
\end{figure}

\indent As to obtain a stationary heat flux, the chain is connected
two heat baths at temperature $T_{+}$ and $T_{-}$, respectively.
Fixed boundary conditions are taken $x_{0}=x_{N+1}=0$. The equations
of motion for the central particles ($i=2,3...N-1$) are
\begin{equation}\label{}
    m\ddot{x}_{i}=-\frac{\partial U(x_{i})}{\partial x_{i}}+K(x_{i+1}+x_{i-1}-2x_{i}).
\end{equation}
The equations of motion for $i=1$ and $i=N$ particles are
\begin{equation}\label{}
    m\ddot{x}_{1}=-\frac{\partial U(x_{1})}{\partial x_{1}}+K(x_{2}-2x_{1})-\gamma\dot{x_{1}}+\xi_{-}(t),
\end{equation}
\begin{equation}\label{}
    m\ddot{x}_{N}=-\frac{\partial U(x_{N})}{\partial x_{N}}+K(x_{N-1}-2x_{N})-\gamma\dot{x}_{N}+\xi_{+}(t),
\end{equation}
where $\gamma$ is friction coefficient and the noise terms
$\xi_{\pm}(t)$ satisfy the fluctuation dissipation relations
$\langle \xi_{-}(t)\xi_{-}(t^{'})\rangle=2\gamma k_{B}T_{-}
\delta(t-t^{'})$, $\langle \xi_{+}(t)\xi_{+}(t^{'})\rangle=2\gamma
k_{B}T_{+} \delta(t-t^{'})$, $k_{B}$ being Boltzmann's constant. The
dot stands for the derivative with respect to time $t$.

\indent For simplicity we set the mass of the particles, the
friction coefficient $m=\gamma=1$. The local heat flux is defined by
$J_{i}=k\langle \dot{x_{i}}(x_{i}-x_{i-1})\rangle$ and the local
temperature is defined as $T_{i}=\langle m\dot{x_{i}}^{2}\rangle$.
$\langle...\rangle$ denotes an ensemble average over time. After the
system reaches a stationary state, $J_{i}$ is independent of site
position $i$, so that the flux can be denoted as $J$.  Thermal
conductivity is evaluated as
\begin{equation}\label{}
    \kappa=\frac{NJ}{T_{+}-T_{-}}.
\end{equation}
 Thermal conductivity represents an effective transport
coefficient that includes both boundary and bulk resistances.

\indent In our simulations, the equations of motion are integrated by using a
second order Stochastic Runge-Kutta algorithm\cite{b2} with a small
time step. Due to the deformation of the potential, the time step
must be less than $10^{-4}$ for large values of $r$.  The
simulations are performed long enough to allow the system to reach a
nonequilibrium steady state in which the local heat flux is a constant
along the chain. To obtain a steady state, the total integration is
typically $10^{8}$ time units. We have checked that this is
sufficient for the system to reach a steady state since the
temperature profile in the central region is linear and the local
heat flux is independent of the site.
\begin{figure}[htbp]
\begin{center}\includegraphics[width=10cm,height=8cm]{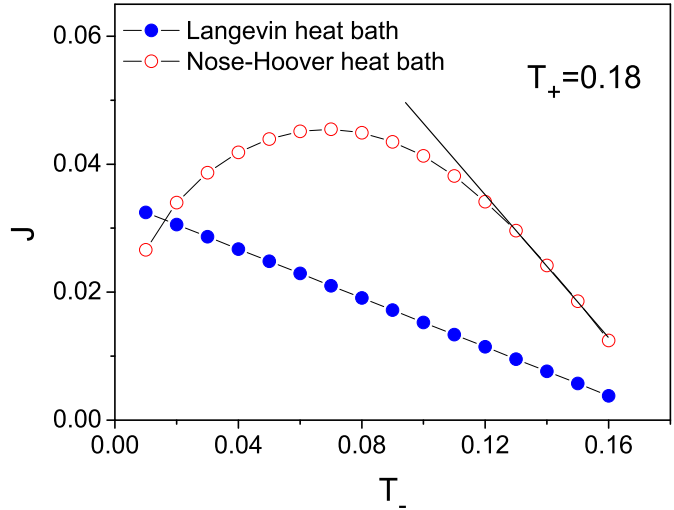}
\caption{(Color online) Heat flux $J$ as function of $T_{-}$ in pure harmonic model for different heat baths at $T_{+}=0.18$.
For pure harmonic lattice, $U(x)=0$ in Eq. (1) and $J\propto T_{+}-T_{-}$. The other parameters are $N=64$ and $K=1.0$.
}\label{1}
\end{center}
\end{figure}

\indent We use Langevin heat baths instead of Nose-Hoover heat baths for the simulations because the use of
Nose-Hoover heat baths might lead to unreliable results in the
nonlinear response regime, particularly at very low or very high
temperatures. The unreliable results are caused by either insufficient equilibration
times or an artifact of using Nose-Hoover thermostats \cite{b3}.  The pure harmonic lattice is a good model to
check the validity of the thermostats. As we known, the heat flux in harmonic model increases linearly with the
temperature difference, $J\propto T_{+}-T_{-}$. However, from Fig. 2 we can see that the heat flux $J$ as a function of
temperature difference is nonmonotonic for Nose-Hoover heat baths, even NDTR occurs for large temperature difference.
 Obviously, it is not true. Therefore, we must be careful for using the Nose-Hoover heat baths.
 For Langevin heat baths, the heat flux increases linearly with temperature difference.
 The Langevin heat baths may be more reliable than Nose-Hoover heat baths .

\section {Numerical results and discussion}

\begin{figure}[htbp]
\begin{center}\includegraphics[width=8cm,height=6.5cm]{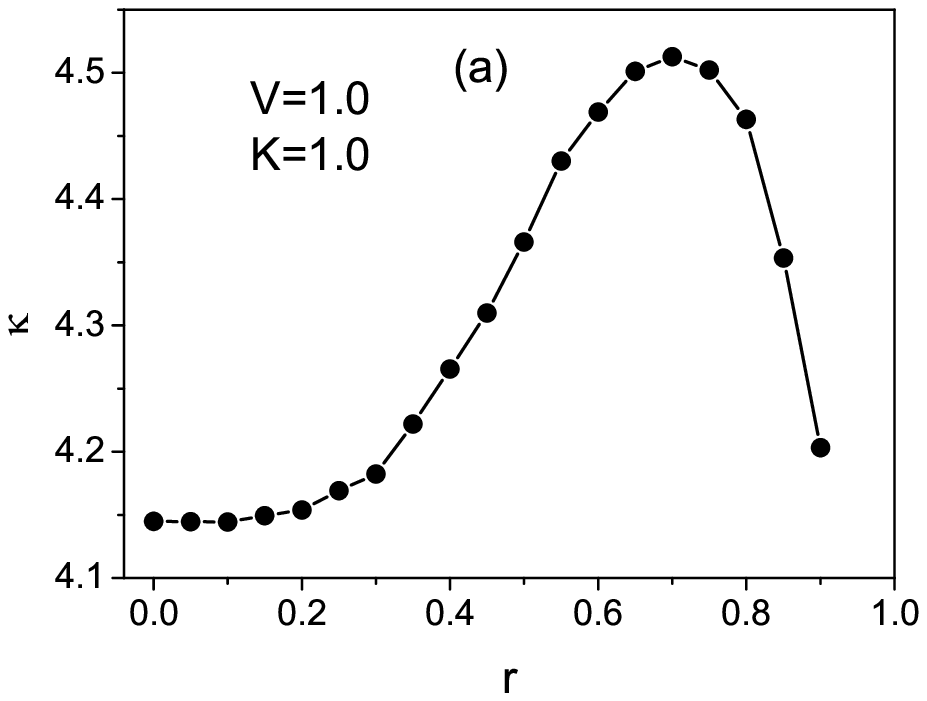}
\includegraphics[width=8cm,height=6.5cm]{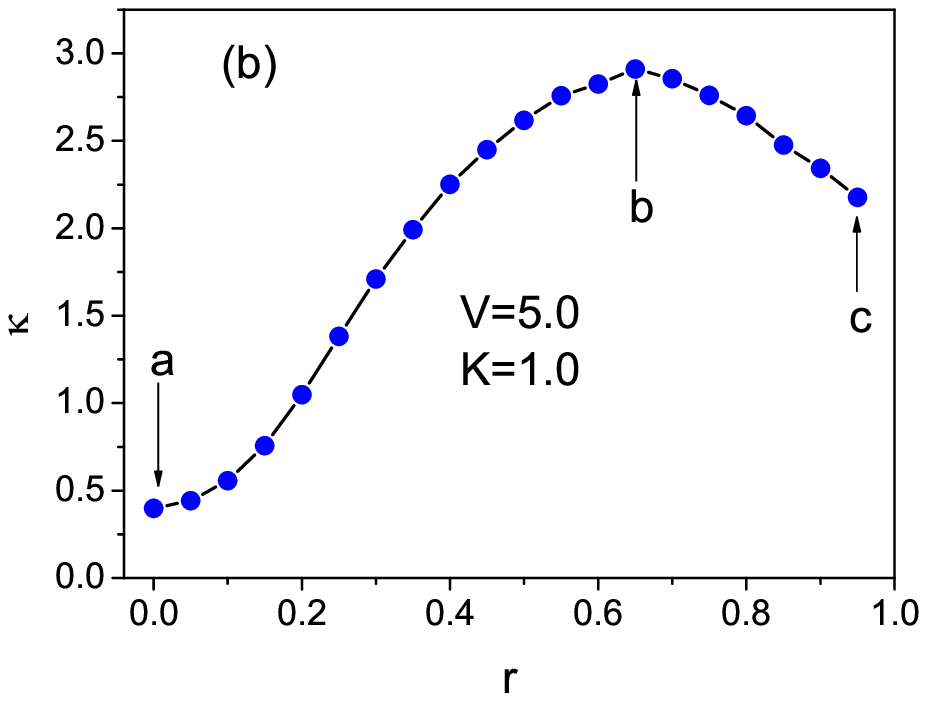}
\includegraphics[width=8cm,height=6.5cm]{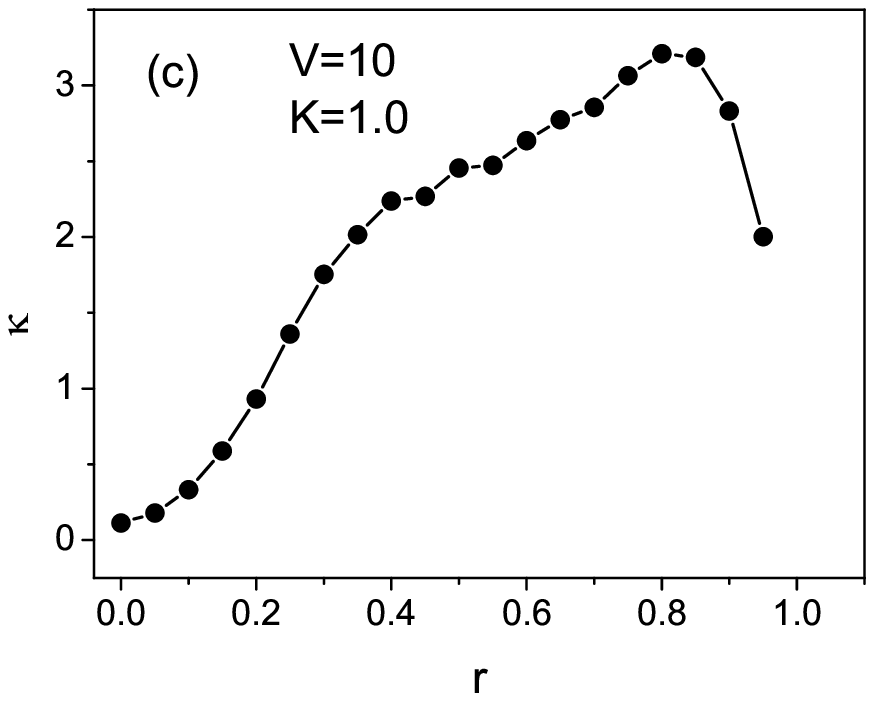}
\includegraphics[width=8cm,height=6.5cm]{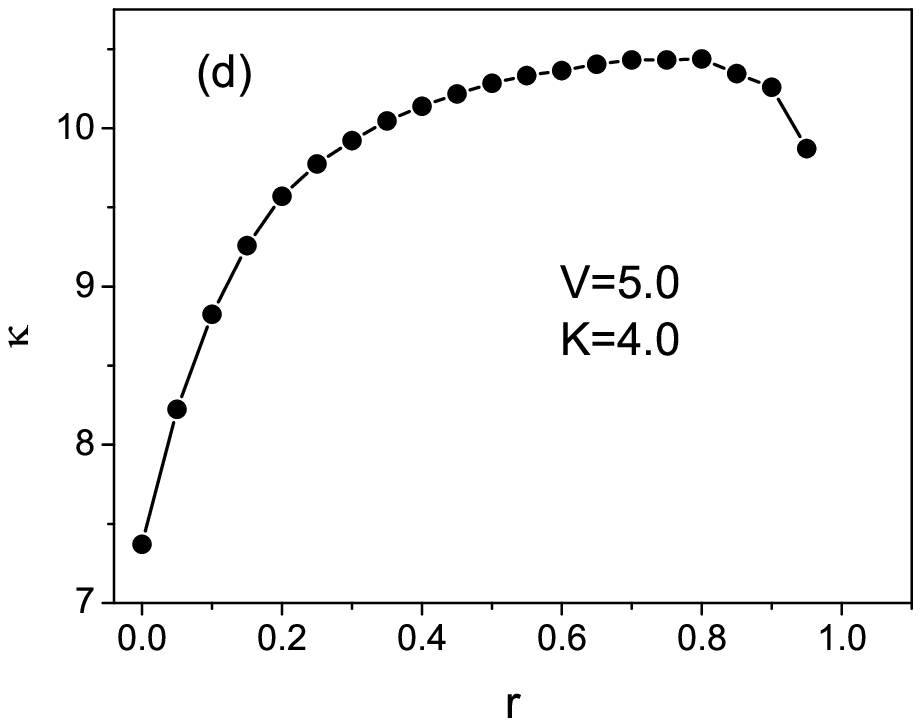}
\includegraphics[width=8cm,height=6.5cm]{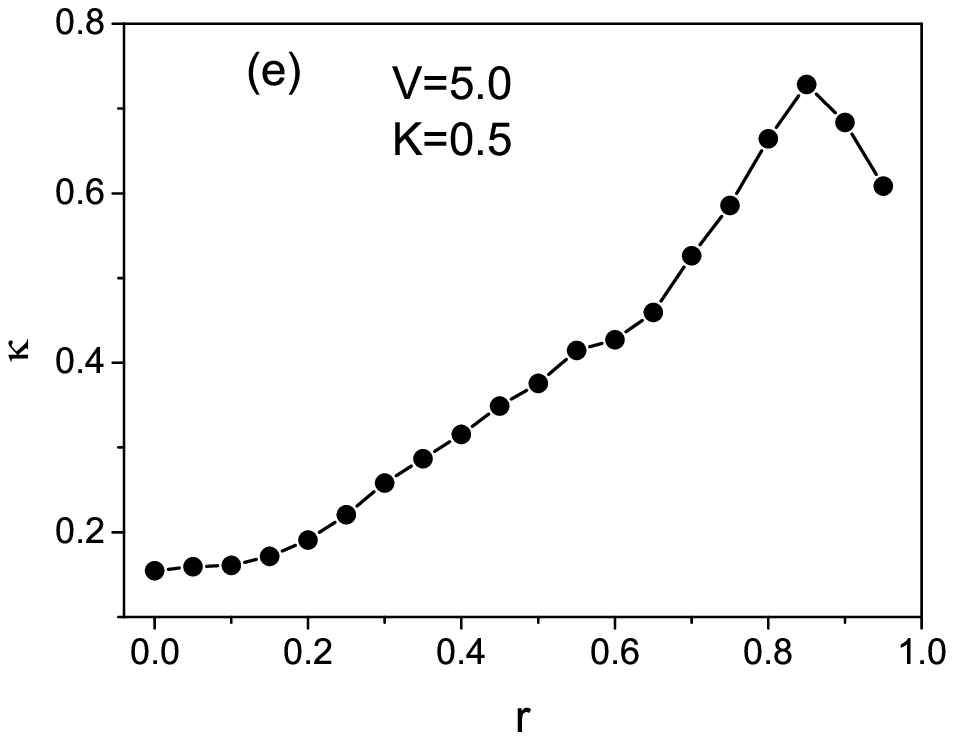}
\includegraphics[width=8cm,height=6.5cm]{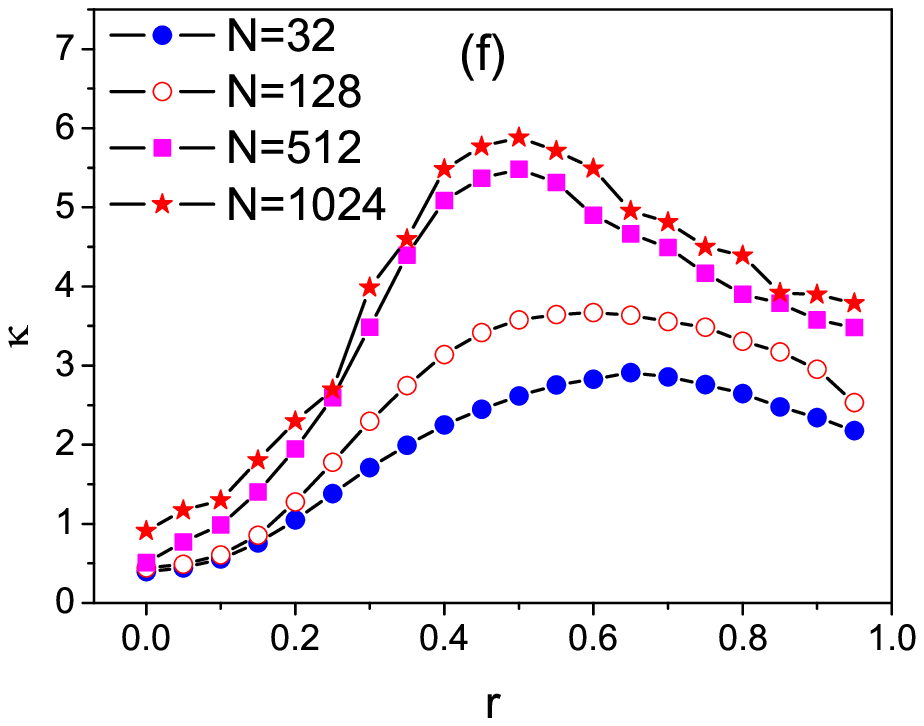}
\caption{(Color online) The dependence of thermal conductivity
$\kappa$ on the shape parameter $r$ for different cases.
(a)$V=1.0$ and $K=1.0$; (b) $V=5.0$ and $K=1.0$; (c) $V=10.0$ and
$K=1.0$; (d) $V=5.0$ and $K=4.0$; (e) $V=5.0$ and $K=0.5$; (f) for
different sizes $N=32$, $128$, $512$, and $1024$ at $V=5.0$,
$K=1.0$. In (a)-(e) $N=32$ and the other parameters are $T_{+}=0.2$
and $T_{-}=0.1$.}\label{1}
\end{center}
\end{figure}

\indent Figure 3 shows thermal conductivity $\kappa$ as a function
of shape parameter $r$ for different cases. From the figure we
can find some interesting and surprising phenomena. The numerical results show that
the deformation can enhance the thermal conductivity greatly.
For example, for case of $V=10$ and $K=1.0$ (shown in Fig. 3 (c)), thermal conductivity can
be increased by about $27$ times ($\kappa=0.113$ and $3.21$ for $r=0.0$ and $0.80$, respectively).
The thermal conductivity $\kappa$ increases at first and then decreases as the
shape parameter $r$ increases. There exists a value of $r$ at
which $\kappa$ takes its maximum.  Now we will give a physical interpret for the novel
phenomenon. From Fig. 1, we can see that the deformable potential
has two inequivalent successive wells with a flat and sharp bottom.
When the particles stay in the flat well, the force acting the
particle is very small, the heat is easy to be transferred from high
temperature heat bath to the low one, resulting in a large thermal
conductivity. Whereas the acting force is very large and the heat
conductivity is small when the particles stay in the sharp well.
When $r$ increases from zero, the probability of the particles in
the flat well is larger than that in the sharp one, the flat well
dominated the heat conduction, so thermal conductivity increases. On
further increasing $r$, the force acting the particles in the sharp
well increases drastically, thermal conductivity is dominated by the
sharp well, so thermal conductivity decreases. Therefore, there is a
peak in the curves. Due to the universe finite size effect in
low-dimensional systems, we consider more system sizes $N=128$,
$512$, and $1024$, as shown in Fig. 3(f), thermal conductivity as a
function of the shape parameter mentioned above is still
invariant. Thermal conductivity increases to a saturation value when
the system size increases since thermal conductivity of the FK model
is finite in the thermodynamic limit. This indicates that these
behaviors induced by the deformation of the potential are not the
small-size effect.  If the phenomenon disappears as the system size increases,
the phenomenon is a small-size effect.

\begin{figure}[htbp]
\begin{center}\includegraphics[width=10cm,height=8cm]{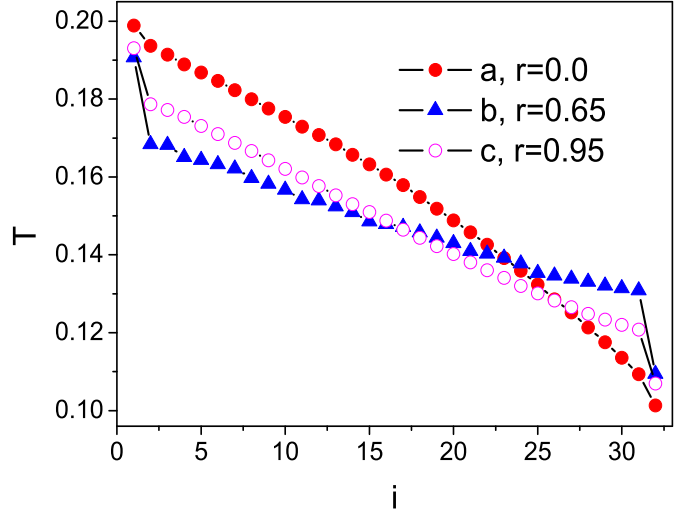}
\caption{(Color online) Temperature profiles for different values of $r$ described
in Fig. 3 (b). }\label{1}
\end{center}
\end{figure}

\indent Figure 4 shows the temperature profiles for $a$, $b$, and
$c$ points shown in Fig. 3 (b).  From the transition of the thermal
transport mode  we can also interpret the different thermal
transport properties. Diffusive and ballistic regions are,
respectively, corresponding to large and small temperature
differences in the middle of the chain.  The thermal conductivity can also be given by
the Debye formula \cite{a1}
\begin{equation}\label{}
    \kappa=\frac{c}{2\pi}\int_{0}^{2\pi}\upsilon(k)l(k)dk,
\end{equation}
where $c$ is the specific heat, $\upsilon(k)$ the velocity of the effective phonon, and $l(k)$
the mean-free path of the effective phonon. For ballistic region, both the velocity $\upsilon(k)$ and the mean-free path $l(k)$ are large,
 so the thermal conductivity is large. When the system goes into the diffusive region, both the velocity $\upsilon(k)$ and
 the mean-free path $l(k)$ decreases, therefore, the thermal conductivity becomes small. It is easy to know that the flat well induces
ballistic transport, whereas the sharp well contributes to diffusive
transport.  When the shape parameter $r$ increases from $r=0$,
the temperature difference in the middle of the chain decreases (from $a$ to $b$),
which indicates that thermal transport mode exhibits a transition
from the diffusive to the ballistic transport. Ballistic transport
means less collision of phonon and then the thermal current will
increase. On further increasing $r$, the temperature difference
increases (from $b$ to $c$), the system tends to diffusive transport and the thermal
current decreases. The competition between the flat well (positive
effect) and the sharp well (negative effect) determines thermal
conductivity.

\begin{figure}[htbp]
\begin{center}\includegraphics[width=10cm,height=8cm]{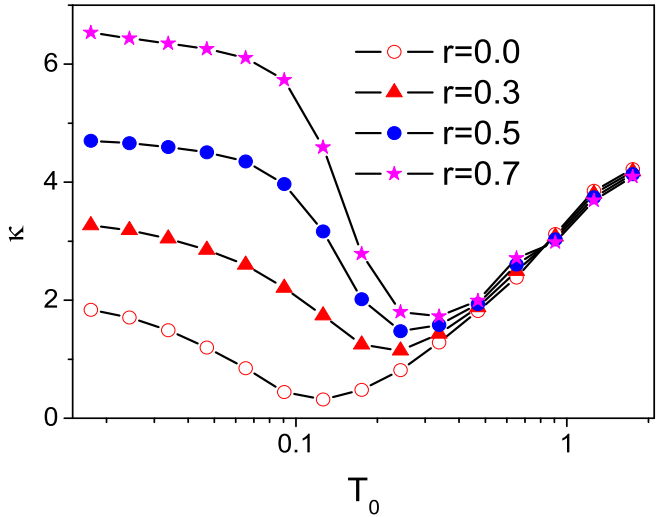}
\caption{(Color online) Temperature dependence of thermal
conductivity $\kappa$ for different values of $r=0.0$, $0.3$, $0.5$,
and $0.7$. Here $V=5.0$, $K=1.0$, $\Delta T=0.02$, and $N=32$.
$T_{0}=\frac{T_{+}+T_{-}}{2}$, $\Delta T=T_{+}-T_{-}$.}.\label{1}
\end{center}
\end{figure}
\indent Figure 5 depicts the dependence of thermal conductivity on
temperature for different values of $r$. Similar to the case of
$r=0$, thermal conductivity in deformable potential has a minimum at
$T_{c}$. Thermal conductivity $\kappa$ decreases monotonically with
$T_{0}$ for $T_{0}<T_{c}$ and increases monotonically with $T_{0}$
for $T_{0}>T_{c}$.  In FK model, there exist low and high temperature limits.
For high temperature limit, $k_{B}T\gg \frac{V}{(2\pi)^{2}}\approx 0.13$, where the substrate potential
is irrelevant, kinks do not play a role, and the FK chain behaves as a weakly interacting phonon gas.
For low temperature limit, $k_{B}T\ll \frac{V}{(2\pi)^{2}}$, where the substrate potential is relevant
and kinks play a role. For low temperatures, the deformation of the
potential affects thermal conductivity drastically. However, the
thermal conductivity becomes independent of the deformation for very
high temperatures. This is because the effect of the substrate
potential disappears for high temperatures, each particle can
jump freely between the minima of the substrate potential. The critical temperature $T_{c}$ is dominated by the sharp well of the potential.
The maximal force from the deformable potential is larger than that from the standard potential. The maximal force from the substrate potential increases with $r$.
It needs higher temperature for all particles jumping freely between the two wells. Therefore, the critical temperature $T_{c}$ shifts to the high temperatures on increasing $r$.

\begin{figure}[htbp]
\begin{center}\includegraphics[width=8cm,height=8cm]{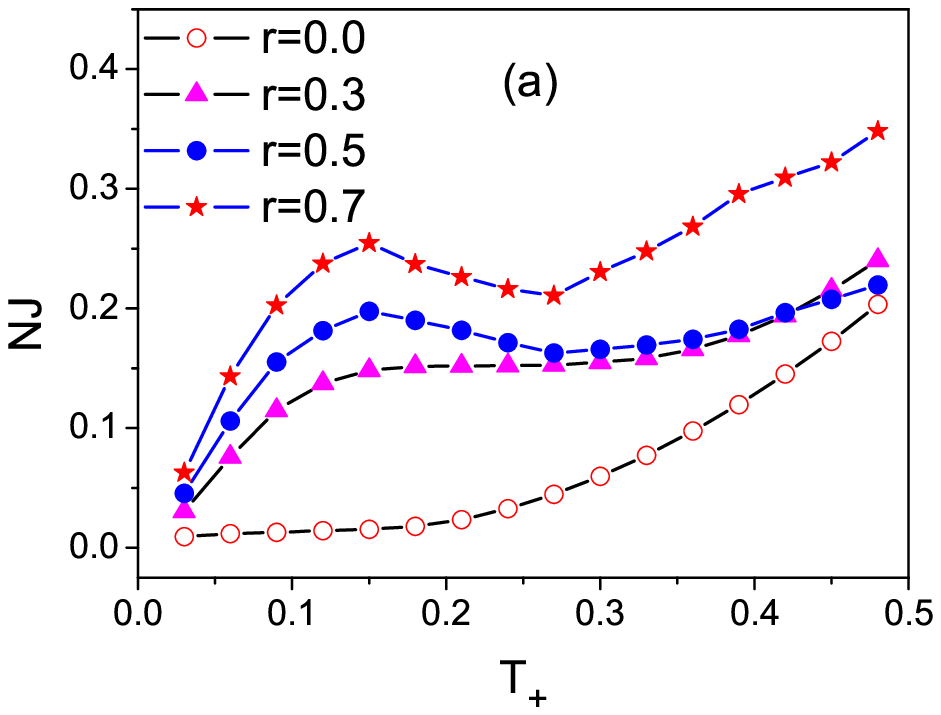}
\includegraphics[width=8cm,height=8cm]{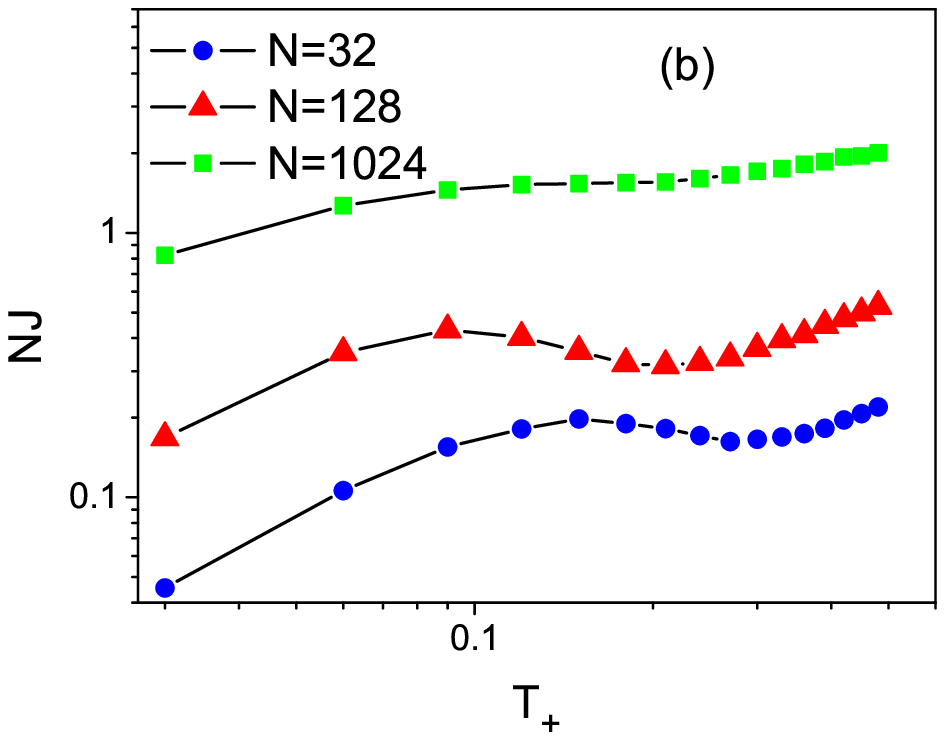}
\includegraphics[width=8cm,height=8cm]{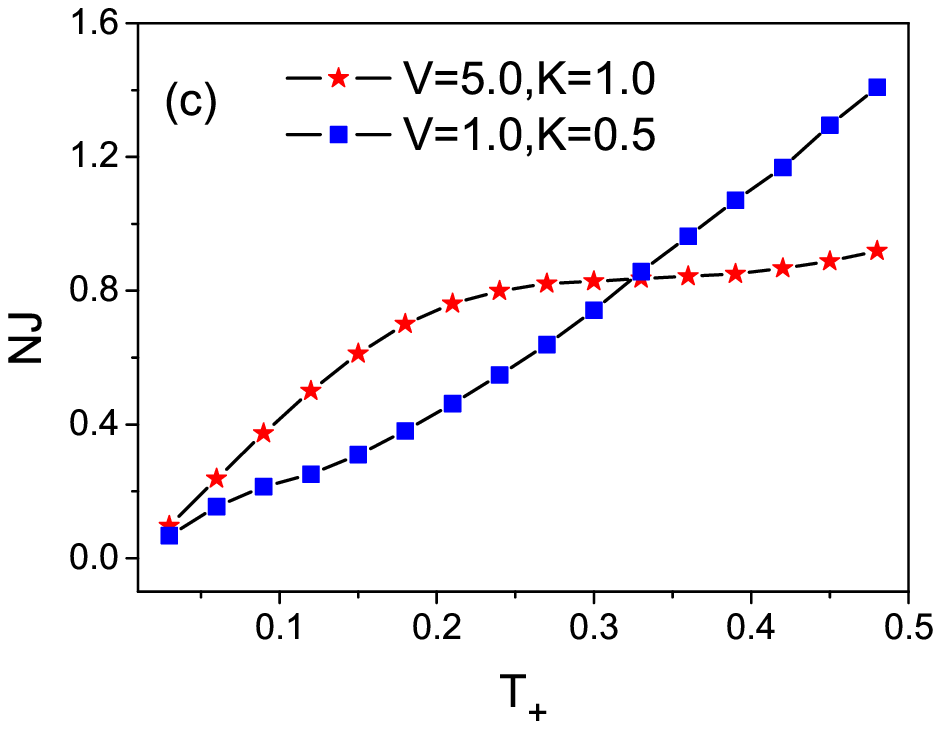}
\caption{(Color online) Total heat flux $J$ as a function of $T_{+}$
for fixing $T_{-}=0.01$. (a) For different values of $r$ at $V=5.0$,
$K=0.5$, and $N=32$; (b) size effects on NDTR at $V=5.0$, $K=0.5$,
and $r=0.5$; (c) for different cases of ($V=5.0$, $K=1.0$) and
($V=1.0$, $K=0.5$) at $N=32$ and $r=0.5$.}\label{1}
\end{center}
\end{figure}

\indent  Now, we return to the phenomenon of NDTR, which refers to
the phenomenon where the resulting heat flux decreases as the
applied temperature difference(or gradient) increases. Figure 6 (a)
shows the total heat flux $NJ$ as a function of $T_{+}$ for regular ($r=0.0$)
and deformable ($r=0.3$, $0.5$, $0.7$) FK chain, respectively, at
$V=5.0$, $K=0.5$, $T_{-}=0.01$, and $N=32$. In our previous work \cite{c1}, for regular FK chain, NDTR can
occur for very low temperature $T_{-}=0.001$. However, for $T_{-}=0.01$, $NJ$ increases with $T_{+}$ monotonically and no NDTR occurs.
Interestingly, on increasing the shape parameter $r$, the
system enters a nonlinear response regime where NDTR occurs.
Therefore, the deformation of the potential can facilitate the appearance of NDTR. Figure
6 (b) shows the size effects on NDTR. The numerical simulations show
the exhibition of NDTR for a system size of $N=32$ and $128$ but not
for the case of $N=1024$. This indicates that the regime of NDTR
becomes smaller as the system size increases, and eventually
vanishes in the thermodynamic limit. From Fig. 6 (c), we can see
that the NDTR will disappear for decreasing $V$ or increasing $K$.
It becomes easier for the particles to overcome the on-site
potential via their thermal energy when $V$ decreases or $K$
increases. Therefore, the system is approaching the harmonic limit
where NDTR cannot occur. In general, it was found that NDTR may
occur if there is nonlinearity in the on-site potential of the
lattice model. However, in a deformable on-site potential, NDTR will
occur in the larger range of the parameters.

\begin{figure}[htbp]
\begin{center}\includegraphics[width=8cm,height=8cm]{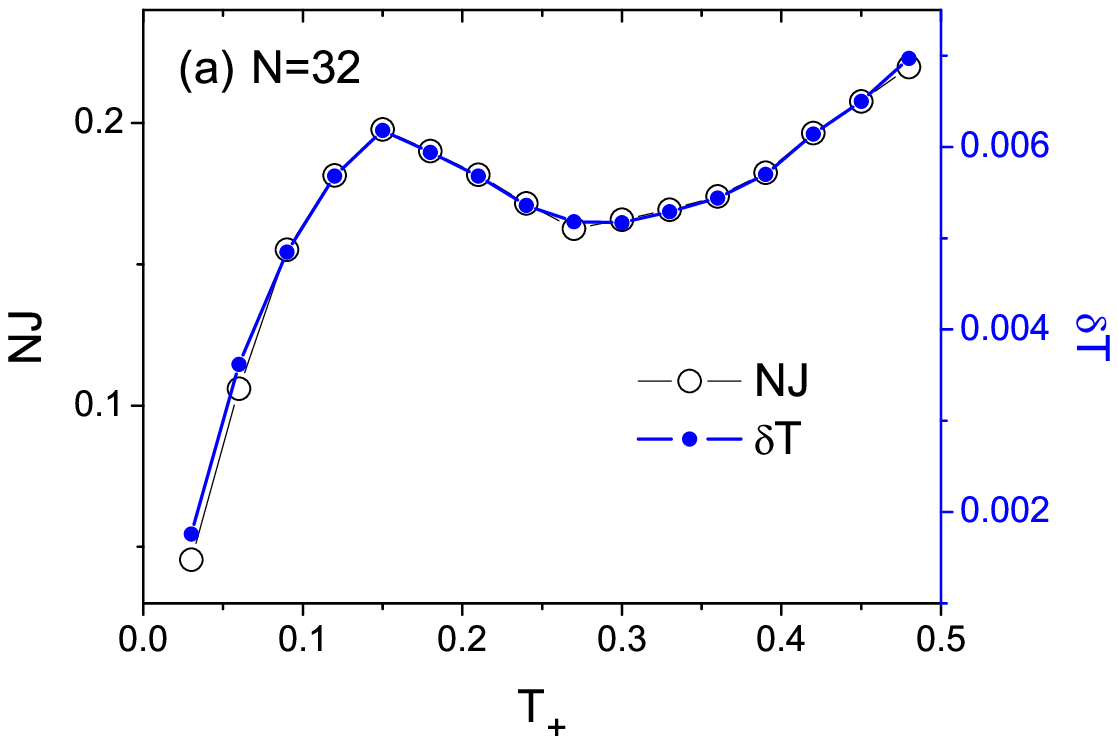}
\includegraphics[width=8cm,height=8cm]{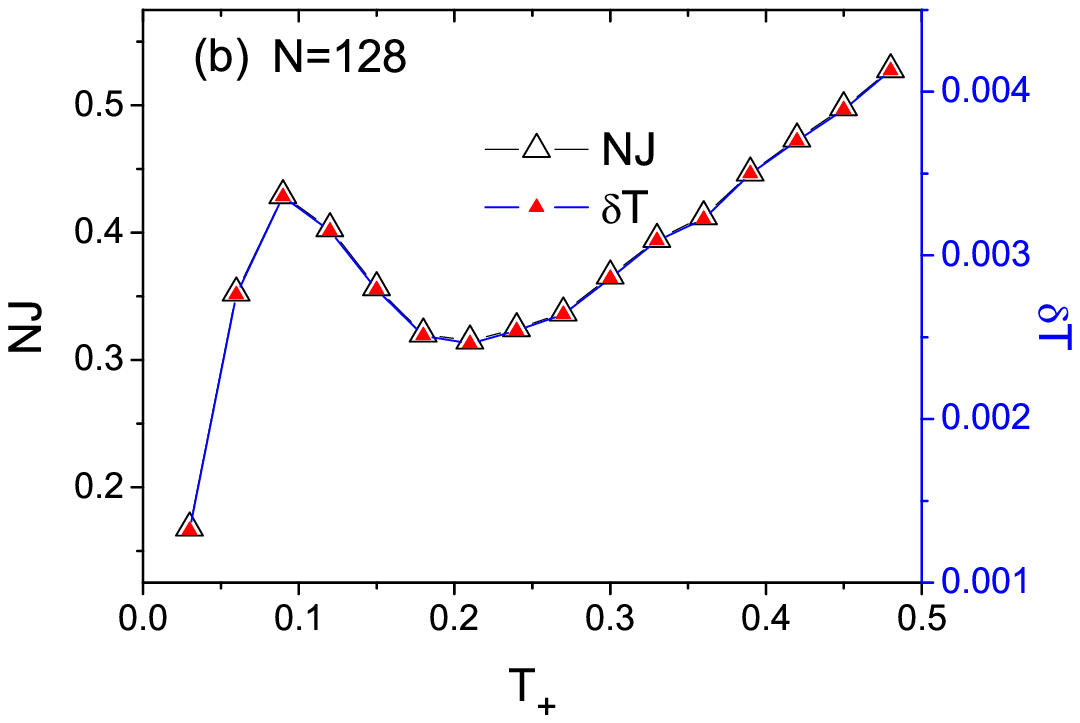}
\includegraphics[width=8cm,height=8cm]{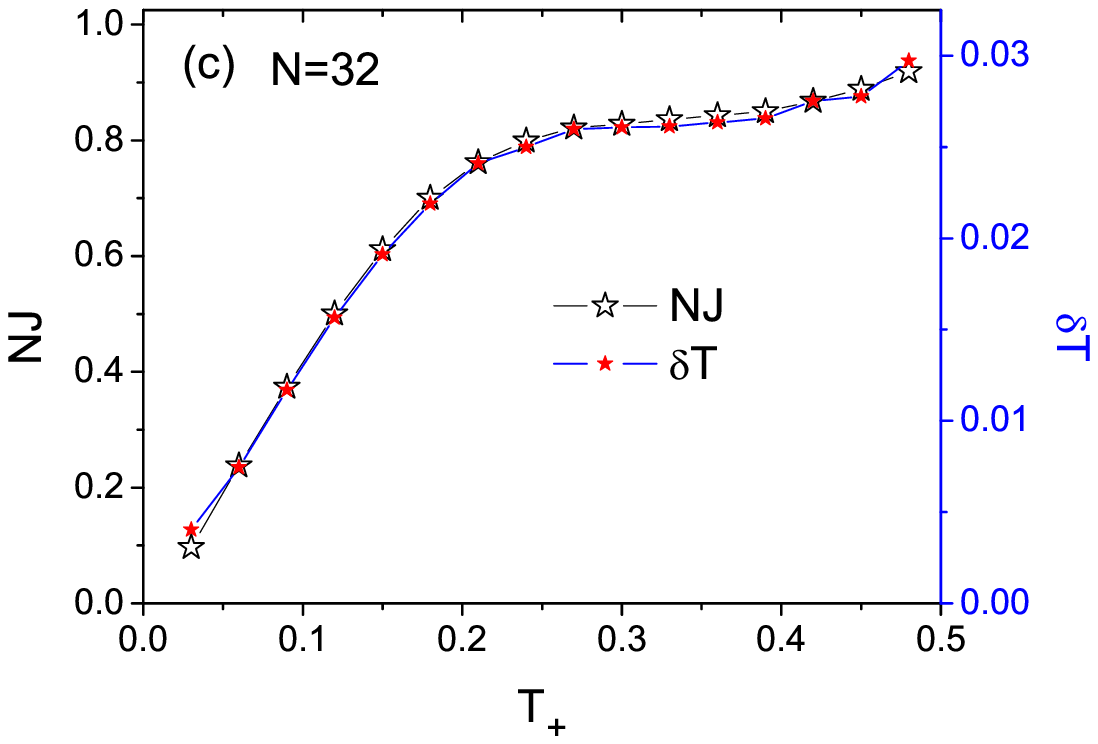}
\caption{(Color online) (a) Total heat flux $NJ$ and the
corresponding boundary temperature jump $\delta T$ as a function of
$T_{+}$ at $r=0.5$. (a) $K=0.5$ and $N=32$; (b)$K=0.5$ and $N=128$;
(c) $K=1.0$ and $N=32$. The other parameters are $V=5.0$,
$T_{-}=0.01$, and  $\delta T\equiv T(N)-T_{-}$.}\label{1}
\end{center}
\end{figure}

\indent As we know, thermal boundary resistance is more important in
small-size systems. From the above results, we can find that NDTR
also occur in small-size systems. Therefore, one can speculate that NDTR may be the result of some kind of boundary mechanisms.  Figure 7
shows how the total heat flux $NJ$ and the boundary temperature jump
\cite{c2} $\delta T\equiv T(N)-T_{-}$ vary with $T_{+}$. From the
figure, we can see that total heat flux $NJ$ exhibits the same
relation with the boundary temperature jump $\delta T$ for different
cases. Therefore, NDTR may be caused by some boundary effects, for
example, the phonon-boundary scattering or thermal boundary
resistance. However, the general physical mechanism of NDTR for both
two segment asymmetric FK chain \cite{a4} and the single chain
described in this paper is not available. Especially, the necessary
and sufficient conditions for NDTR are still lacking.

\indent For convenience of numerical calculations, all physical quantities are dimensionless.
However, it is necessary to give units of measurement for the potential experiments. First, it can
give us very useful information about the corresponding true temperature to that one we used and enable us to gain some
physical insights. The real temperature $T_{r}$ is related to $T$ through the relation \cite{a7,c3}
\begin{equation}\label{}
    T_{r}=\frac{m\omega_{0}^{2}L^{2}}{k_{B}}T,
\end{equation}
where $m$ is mass of the particle, $\omega_{0}=\sqrt{K/m}$ is the oscillating frequency, $L$ is the period of external
potential, and $k_{B}$ is the Boltzmann constant. For the typical values of atoms, we have
\begin{equation}\label{}
    m\sim 10^{-26}-10^{-27} kg, \indent
    k_{B}=1.38\times 10^{-23} JK^{-1},\indent
    L\sim 10^{-10}m,\indent
    \omega_{0} \sim 10^{13} sec^{-1}.
\end{equation}
\indent From Eqs. (8) and (9), we can find that $T_{r}\sim (10^{2}-10^{3}) T$, which means that the room temperature corresponds to the
dimensionless temperature $T$ about the order of $0.1-1.0$.

\section{Concluding Remarks}
\indent In summary, heat conduction
through the deformable FK lattices is extensively studied by using nonequilibrium molecular
dynamics simulations. It is found that the deformation of the substrate
potential induces the important effects on thermal conductivity and
NDTR. Thermal conductivity can be enhanced greatly by changing the
shape parameter. Due to the competition between the sharp well and the flat well of the substrate potential,
there exists an optimized value of $r$ at which the thermal conductivity $\kappa$ takes its maximal value. \
This behavior does not disappear as the system size varies. This means that the phenomenon is not a small-size effect, which is crucial for
possible practical application as to control the heat flux. Thermal
conductivity is drastically affected by the deformation for low
temperatures, whereas the effect of the deformation will disappear
for high temperatures. Interestingly, NDTR will appear when the
shape parameter is increased, which may be observed and plays
an important role in the operation of thermal devices
\cite{a4,a5,a6}. It is easier to obtain the phenomenon of NDTR in
the deformable FK lattices. The range of NDTR becomes smaller for
increasing $K$ or decreasing $V$.  The phenomenon of NDTR will
disappears in the thermodynamic limit ($N\rightarrow \infty$) and it is a small-size effect.
We believe that the introduction of deformable FK model for heat conduction brings many
interesting new features for physical applications, such as heat
conduction in lattice with diatomic basis, dual lattices,  or crystals with dislocations.

\indent We would like to thank members of the Centre for Nonlinear
Studies for useful discussions.  This work was supported in part by
Hong Kong Research Grants Council (RGC), the Hong Kong Baptist
University Faculty Research Grant (FRG), National Natural Science
Foundation of China (Grant No. 30600122), and GuangDong Provincial
Natural Science Foundation (Grant No. 06025073).

\end{document}